\documentclass[twocolumn]{aastex62}
\usepackage{verbatim, graphicx,  ifthen, pbox}
\usepackage{eso-pic}
\usepackage{CJKutf8}
\usepackage{xcolor}
\usepackage{hyperref}
\usepackage{breakurl}
\usepackage{amsmath}
\usepackage{graphicx}
\usepackage{verbatim}
\usepackage{booktabs}
\usepackage[T1]{fontenc}
\usepackage{ae,aecompl}

\usepackage{newtxtext,newtxmath}
\renewcommand{\vec}[1]{\mathbf{#1}}

\begin{document}
\title{The Resonance Hopping Effect in the Neptune-Planet Nine System}

% Don't change these lines

\author[0000-0001-7721-6457]{T.~Khain}
\affiliation{Department of Physics, University of Michigan, Ann Arbor, MI 48109, USA}
\author[0000-0002-7733-4522]{J.~C.~Becker}
\affiliation{Division of Geological and Planetary Sciences, California Institute of Technology, Pasadena, CA 91125, USA}
\affiliation{Department of Astronomy, University of Michigan, Ann Arbor, MI 48109, USA}
\author[0000-0002-8167-1767]{F.~C.~Adams}
\affiliation{Department of Physics, University of Michigan, Ann Arbor, MI 48109, USA}
\affiliation{Department of Astronomy, University of Michigan, Ann Arbor, MI 48109, USA}

%\correspondingauthor{Tali Khain}

\begin{abstract}
The observed physical clustering of the orbits of small bodies in the distant Kuiper Belt (TNOs) has recently prompted the  prediction of an additional planet in the outer solar system. Since the initial posing of the hypothesis, the effects of Planet Nine on the dynamics of the main cluster of TNOs -- the objects anti-aligned with its orbit -- have been well-studied. In particular, numerical simulations have revealed a fascinating phenomenon, referred to as ``resonance hopping", in which these objects abruptly transition between different mean-motion commensurabilities with Planet Nine. In this work, we explore this effect in greater detail, with the goal of understanding what mechanism prompts the hopping events to occur. In the process, we elucidate the often underestimated role of Neptune scattering interactions, which leads to diffusion in the semi-major axes of these distant TNOs. In addition, we demonstrate that although some resonant interactions with Planet Nine do occur, the anti-aligned objects are able to survive without the resonances, confirming that the dynamics of the TNOs are predominantly driven by secular, rather than resonant, interactions with Planet Nine.\\
\end{abstract}

\section{Introduction} \label{sec:intro}

In the past few years, the outer solar system community has exploded with hundreds of papers exploring the Planet Nine hypothesis, the prediction of an additional planet in our solar system \citep{2014Natur.507..471T, 2016AJ....151...22B, P9Review}. The basis of this prediction is the observation that the orbits of the most distant, extreme trans-Neptunian objects (ETNOs; in this work, this definition includes objects whose perihelion distances are greater than Neptune's semi-major axis and whose semi-major axes are greater than 250 AU) tend to point in the same direction in physical space (i.e., they are both apsidally confined and clustered about a common orbital plane) in a statistically significant way {(see discussion in \citet{P9Review})}. This apparent asymmetry, if real\footnote{Note that some papers argue that the asymmetry is a symptom of observational bias rather than a physical reality (see discussions of both sides of this issue in \citealt{Shankman2017,Brown2017,Brown2019,Kaib2019,Kavelaars2019}).}, cannot be explained by the current eight-planet solar system. Even factors external to the solar system such as the galactic tide or stellar fly-bys do not appear to reproduce the observed asymmetry \citep{Clement2020}. Unless the outer solar system includes significantly more objects {exterior to Neptune's orbit} %a much more massive (and distant) scattered disk 
than observations seems to imply \citep{Madigan2016MNRAS,Sefilian2019AJ}, these observations indicate an additional planetary-mass object, generally known as ``Planet Nine''. This planet would reside on a distant, eccentric orbit that is aligned in the direction opposite to the main cluster of the so-called anti-aligned ETNOs (where anti-aligned objects would be that fall within a population centered at $\delta\varpi \sim 180$ with respect to Planet Nine).
In addition to demonstrating that Planet Nine can create the observed orbital confinement, \citet{2016AJ....151...22B} predicted that Planet Nine would generate a population of distant, high-inclination objects. The recent discovery of the first object in this class, the ETNO 2015 BP$_{519}$ ($a = 450$ AU, $e = 0.92$, and $i = 54^{\circ}$), serves as an independent line of evidence for the existence of Planet Nine \citep{2018AJ....156...81B}. %High-inclination orbits with these characteristics are readily produced in solar systems containing Planet Nine, but would require an alternate explanation in its absence. 

Recent literature on the hypothesis includes discussions of the effect of Planet Nine on the dynamics of the TNOs, the relevance of observational bias on the prediction of the planet, its possible formation mechanisms, and more, and has been summarized in the review \citet{P9Review}. The stability of the anti-aligned population and its dependence on perihelion distance was explored in \citet{summerwork}, in which a large set of numerical simulations were used to evaluate the properties of different KBO populations that would be generated by Planet Nine. However, the dynamical behavior of these objects with varying perihelia distances is not yet fully understood. In particular, an important aspect of the dynamics that is still under debate is whether the apsidal alignment of the ETNOs occurs through a dominantly secular mechanism or primarily through resonant interactions with Planet Nine.

In their original paper, \citet{2016AJ....151...22B} present evidence from numerical simulations that suggest the ETNOs reside in mean-motion resonances with Planet Nine. In contrast, \citet{2016A&A...590L...2B} show that secular interactions are sufficient to cause the apsidal alignment of the TNOs. Using a full numerical study, however, \citet{2017AJ....154...61B} evaluate the dynamical stability of a number of observed ETNOs and find that many experience an effect referred to as ``resonance hopping", in which objects jump from one mean-motion commensurability with Planet Nine to another. In further work, \citet{2017AJ....154..229B} expand the theoretical analysis to include secular and semi-analytical resonant contributions and show that both components can create the observed anti-alignment. 

Even though secular effects are sufficient to generate the apsidal alignment, this finding does not guarantee stability for the ETNOs. \citet{2017AJ....154..229B} point out that since the anti-aligned ETNOs lie in orbit-crossing configurations with Planet Nine (assuming the orbits reside in the same plane), they must reside in mean-motion resonances to survive. Such resonant configurations thus allow the ETNOs to avoid close encounters with the planet through the mechanism of phase protection \citep{2002mcma.book.....M}.

Most of the aforementioned studies were conducted using a planar approximation of the real system. In a fully three-dimensional system with a range of inclinations, however, the ETNOs may be able to avoid encounters with the planet even outside of resonance \citep{Li2018}. As a result, the exact balance between the secular and resonant mechanisms in regulating the apsidal alignment and stability of the ETNOs remains an open question. Previous works have considered the problem with either a full three dimensional treatment \citep{2016AJ....151...22B, Khain2018p3, Clement2020}, coplanar treatment \citep{Bromley2016, 2017AJ....154..229B, summerwork}, or with a near coplanar treatment \citep[3 degrees in][]{Li2018}. 

In this present work, our goal is to provide a coherent picture of the dynamics of the anti-aligned TNOs in the presence of both Neptune and Planet Nine. In Section \ref{sec:sims}, we consider the two main dynamical modes of the anti-aligned TNOs found in simplified numerical simulations. In Section \ref{sec:scattering}, we highlight the role played by Neptune in the resonance hopping mechanism by presenting results of a simple theoretical model and comparing its predictions to numerical scattering experiments in a population of test particles. Section \ref{sec:disc} describes how the evolution of a TNO is influenced concurrently by Neptune and Planet Nine. Finally, in Section \ref{sec:conclude}, we present a summary of our results and suggest avenues for future work. 

\section{Numerical Simulations} \label{sec:sims}

We begin by running a suite of numerical simulations that aim to model the evolution of the outer solar system in the presence of Planet Nine. Following the set-up in \citet{summerwork}, we carry out N-body integrations that span $4$ Gyr of integration time, and utilize the \texttt{mercury6} software package \citep{mercury}, employing the hybrid Wisdom-Holman/Bulirsch-Stoer algorithm. 

We consider a simplified model of the solar system, which includes Neptune and Planet Nine as active bodies, but replaces Jupiter, Saturn, and Uranus with a solar $J_2$ moment as in \citet{2016AJ....151...22B}. That is, we account for the gravitational potential of the three planets by adding a non-zero quadrupolar field to the Sun, and extending its radius to Uranus' semi-major axis. This approximation of the giant planets as a quadrupolar term on the Sun's potential has been used commonly in the literature in order to make the computational problem more tractable \citep[ex:][]{2016AJ....151...22B, Millholland2017, 2017AJ....154..229B, summerwork, 2018AJ....156...81B, Eriksson2018}, although we note that some works insetad use fully active particles for the giant planets \citep[ex: ][]{Marcos2016, Shankman2017, Caceres2018, Clement2020}.
{In all works mentioned above, the TNOs are treated as non-interacting particles, whose orbits are affected by the gravitational influence of the star and planets but not each other. The treatment of TNOs as massless\footnote{Note that if the TNOs are treated as \emph{massive} non-interacting test particles, numerical (non-physical) complications can arise depending on the integration scheme - see \citet{Peng2020} for a discussion of this potentially problematic effect.} test particles represents well the true behavior of TNOs since the outer solar system has a very low density of objects and TNOs are not signficantly pertrubed by other TNOs. }

In {our} semi-averaged model, we introduce a non-inclined 10 $m_{\oplus}$ Planet Nine with semi-major axis $a_9 = 700$ AU, eccentricity $e_9 = 0.6$, and set the timestep to $16$ years (one-tenth of Neptune's orbital period). Although a range of possible Planet Nine orbits have been proposed in the literature, here we focus on understanding the resonance hopping mechanism in the presence of the canonical Planet Nine (see Section \ref{sec:disc} for further discussion). 

%Range of potential P9 orbits are possible, but robust. Here we are just studying a specific mechanism, discussed later.

To model the TNOs in the outer Kuiper belt, we include $\sim$10,000 massless test particles in our simulations. The orbits of these synthetic objects are drawn from the following distributions: perihelion distance is uniform with $q \in [30, 300]$ AU, semi-major axis is uniform with $a \in [150, 1000]$ AU, inclination follows a half-normal distribution centered at $i = 0^\circ$ with $\sigma_i = 5^\circ$, and the argument of perihelion ($\omega$), longitude of ascending node ($\Omega$), and mean anomaly ($M$) are uniformly drawn from the full $[0, 360)$ degree range. All generated objects with $q > a$ are discarded from the simulations; in the allowed regions, the $a$ and $q$ distributions are uniform.

In these numerical simulations, we identify the test particles that experience apsidal anti-alignment with Planet Nine for most of their evolution. To do this, we consider the direction in which the orbit of each TNO points with respect to the orbit of Planet Nine. More specifically, we compute the offset of longitude of perihelion of each TNO from Planet Nine, $\Delta \varpi = \varpi - \varpi_9$. The objects of interest - the anti-aligned objects - are those with $\Delta \varpi$ that librates about $180^{\circ}$. 

In the sections that follow, we study that dynamics of these anti-aligned objects in detail, extending the analysis performed in \citet{summerwork}. In particular, we focus on the evolution of the semi-major axis and its timing with the perihelion distance oscillation cycle.

\subsection{Resonance Hopping} \label{sec:hop}

\begin{figure*}
  \begin{center}
      \leavevmode
\includegraphics[width=170mm]{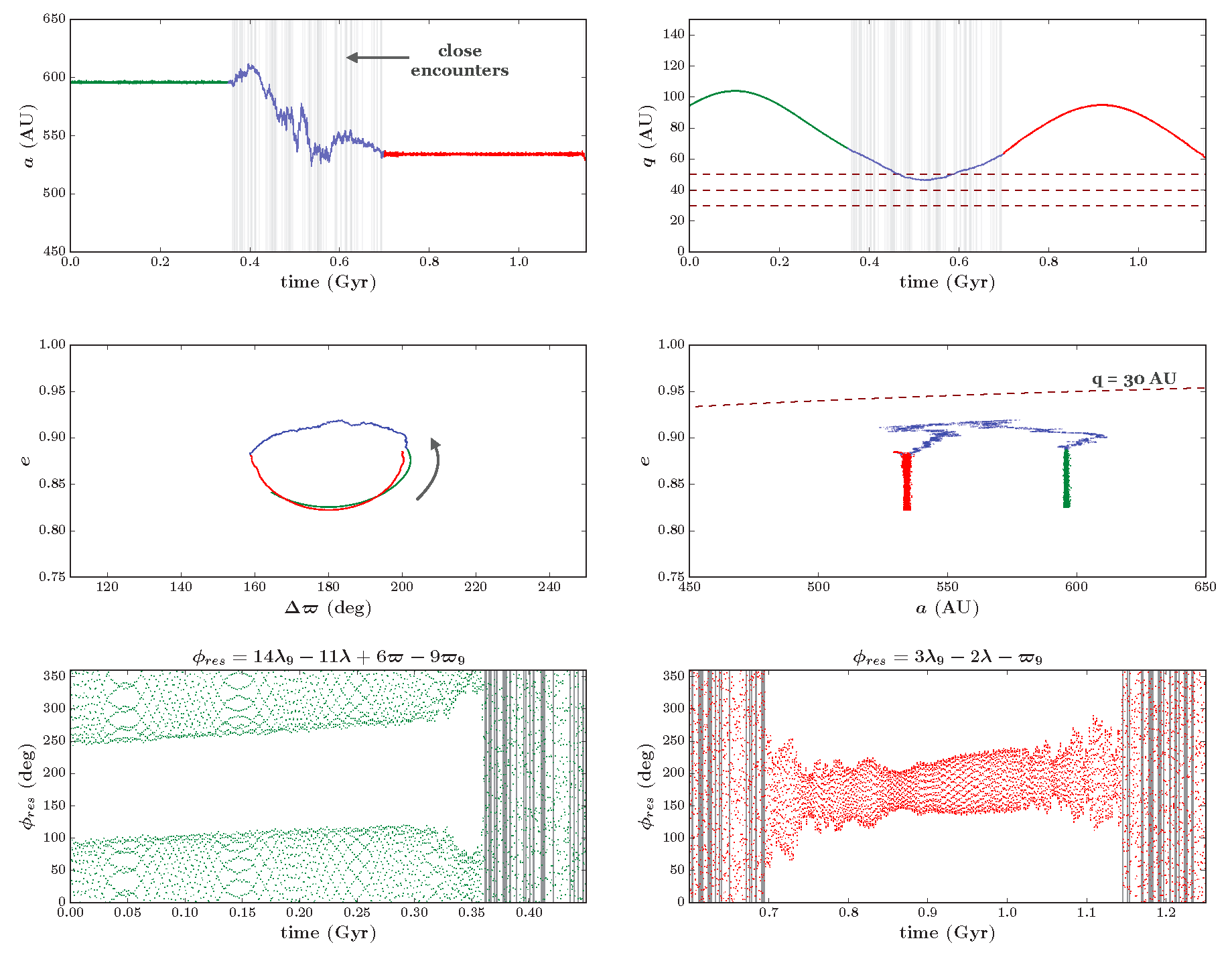}
\caption{An example of a resonance hopping event in the numerical simulations. The top left panel shows the semi-major axis evolution; the two constant $a$ regions, denoted in green and red, correspond to mean-motion resonances with Planet Nine. The associated resonant angles are shown in the bottom two panels. The middle scattering region in the top left panel is the ``hop", and occurs at the bottom of the perihelion distance oscillations shown in the top right panel. The dashed horizontal lines denote $q = 30,\ 40,\ 50$ AU.} The left middle panel confirms that the test particle shown is in fact in an anti-aligned configuration with Planet Nine and experiences apsidal libration. The middle right panel presents an additional visualization of the two regimes: vertical evolution (resonance) and horizontal evolution (scattering). Vertical gray lines throughout the panels show times of close encounters of the test particle with Planet Nine.
\label{fig:hop}
\end{center}
\end{figure*}

In the known outer solar system, the resonant structure of the scattered disk evolves as mediated by interactions with Neptune \citep{Duncan1997}, allowing the trapping of objects into temporary mean motion resonance with Neptune in a process called resonance sticking \citep{Gallardo2006Icar, Lykawka2007}. These objects may spend more time out of resonances than in resonances, only `sticking' in the resonances for relatively short amounts of time, so that the number density of `stuck' objects is dominated by those with short sticking lifetimes. \citep{Yu2018}.

The resonance hopping mechanism, driven by interactions with Planet Nine, was observed by \citet{2017AJ....154...61B} in numerical integrations of the observed extreme TNOs (using their measured orbital elments). In this effect, an object spends a significant amount of time in a mean motion-resonance with Planet Nine; then, it exits the resonance and experiences rapid semi-major axis variation; and finally the object ``hops" into another resonance, where it resides for a relatively long time. This mechanism is related to that of ``nodding'' \citep{nodding2013} where planets initially reside in mean motion resonance, leave the resonant configuration so that the resonant angles circulate for several cycles, and then re-enter the resonance. 

To identify potential resonance hopping events in our simulations, we run the resonance identification algorithm developed in \citet{Khain2020} on the anti-aligned population. We search for resonance arguments that experience bounded oscillations over time, where the angle is given by the d'Alembert relation,
\begin{equation}
\phi_{res} = p\lambda_9 -  q\lambda - r\varpi_9 - s\varpi,
\end{equation}
where $\lambda$ is the mean longitude of the body, $\varpi$ its longitude of perihelion, and $p, q, r$ and $s$ are any integers that satisfy the constraint $p = q + r + s$. This algorithm allows us to find resonant interactions between the test particles and Planet Nine. {We search for resonances that last an appreciable fraction of the simulation lifetime (at least $\sim10$ Myr), and identify all resonances with $p,q < 30$.} 

%An example of such a particle is shown in Figure \ref{fig:hop}. This is an anti-aligned object, as can be seen from the apsidal libration about $\Delta \varpi = 180^{\circ}$ in the middle left panel. That is, during the first billion years of evolution shown, the object is perpetually in an orbit-crossing configuration with Planet Nine. Now, in our analysis, we are looking to study the dynamics of resonant objects. In order to check if this object may be experiencing resonant interactions, let us consider its semi-major axis behavior. 

An example of an anti-aligned particle is shown in Figure \ref{fig:hop}. Due to its apsidal libration about $\Delta \varpi = 180^{\circ}$ (middle left panel), the object is perpetually in an orbit-crossing configuration with Planet Nine. A possible explanation for this object's stability (defined as its continued presence in the solar system without collisions with other bodies or excursions in semi-major axis beyond 10,000 AU) is that this particle is in resonance with Planet Nine. To check if this is the case, let us first consider its semi-major axis behavior.

The top left panel of Figure \ref{fig:hop} shows the semi-major axis time evolution. Note that there are two constant regions, shown in green and in red, and a varying semi-major axis region in the middle in blue. During the two constant regions, the test particle resides interior to Planet Nine in a mean-motion resonance: in the $14$:$11$ resonance for the first 0.35 Gyr, and then in the $3$:$2$ resonance from about 0.7 Gyr to 1.15 Gyr. This behavior is indicated by the bottom two panels, which show the time evolution of the corresponding librating resonance arguments.

The vertical lines in the top two and bottom two panels depict the times of close encounters with Planet Nine (times when the distance between the test particle and Planet Nine is less than Planet Nine's Hill radius, a distance which in this case corresponds to about 6 AU). Close encounters do not occur when the object is in resonance. Due to the phase protection mechanism of the resonance, such particles are instead able to avoid Planet Nine in the orbit crossing regions. In contrast, the close encounters occur frequently when the object is not in resonance, as shown by the blue portions of the evolutionary curves in the figure. 

\begin{figure*}
  \begin{center}
      \leavevmode
\includegraphics[width=170mm]{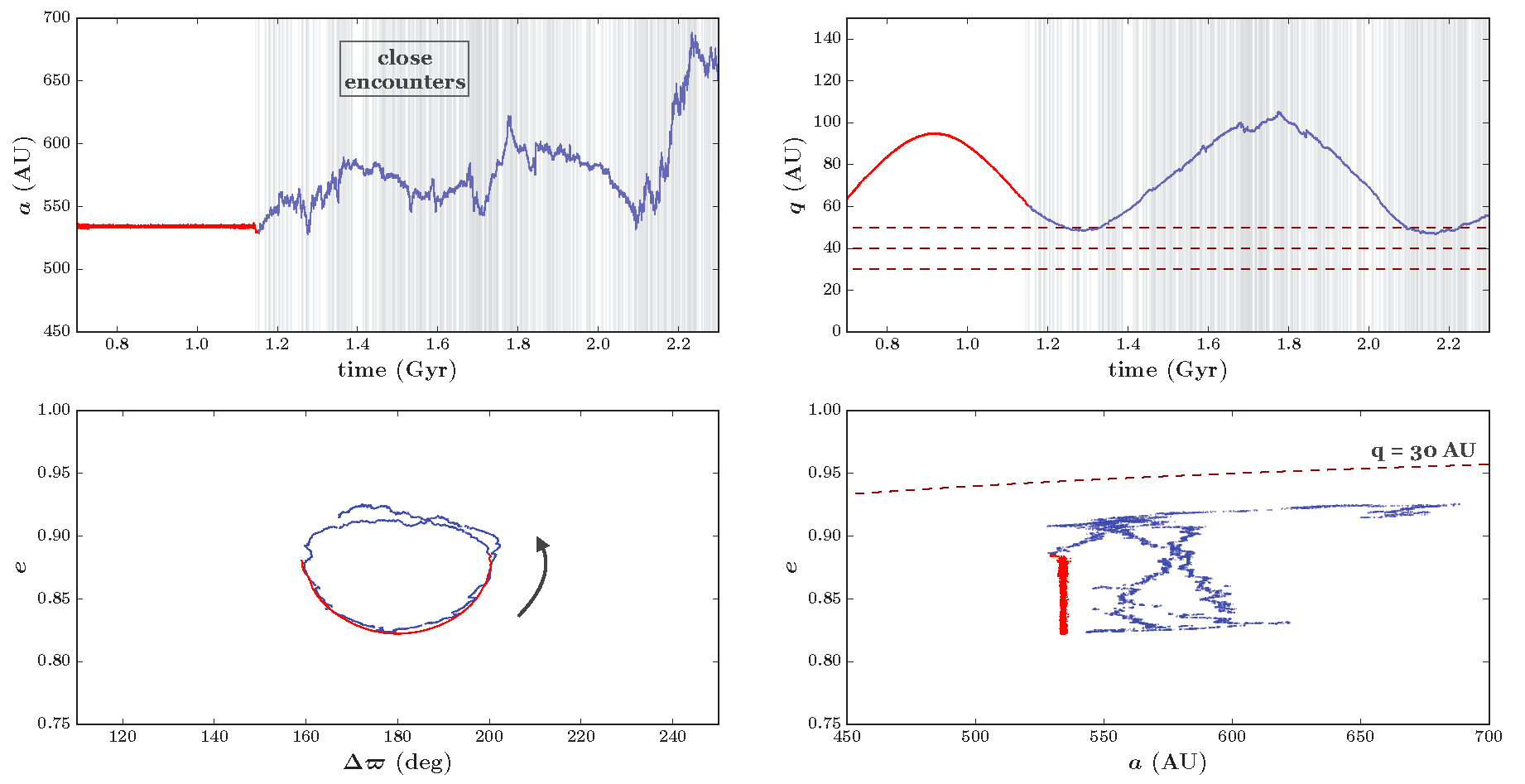}
\caption{The continuation of the evolution of the same test particle as presented in Figure \ref{fig:hop}. Starting out in resonance with Planet Nine, the particle fails to attain a new resonance after experiencing scattering events while at the first plotted local minima of its perihelion distance oscillations. Since the object now resides in an orbit crossing configuration and is not phase protected by a resonance, the ensuing close encounters result in scattering semi-major axis behavior (top left and bottom right panels). The encounters do not significantly affect the perihelion distance oscillations (top right panel) and the apsidal libration (bottom left panel).}
\label{fig:nonres}
\end{center}
\end{figure*}

Now, let us consider the evolution of the perihelion distance, $q$, of the object. Due to secular interactions, the perihelion distance of the anti-aligned objects experiences oscillatory behavior, as shown in the top right panel. By comparing the time intervals for each of the semi-major axis regimes discussed above (note the color-coding), we find that the non-resonant dynamics occurs near the local minimum of the perihelion distance cycle. The object falls out of resonance as the perihelion $q$ decreases, and hops back into resonanace as $q$ increases. Interestingly, the perihelion distance oscillations remain essentially unaffected by the disparate semi-major axis behavior of the particle.

The two regimes are further visualized in the middle right panel, on the $e-a$ plane. The dashed curve across the top of the plot depicts a constant perihelion distance of $30$ AU, which corresponds to the semi-major axis of Neptune's orbit. In this plane, the resonances are easily identifiable as vertical pillars: times when the semi-major axis remains constant. The varying semi-major axis connects the two resonances in the ``hopping" part of the effect. 

Summarizing the evolution of this representative anti-aligned particle, we arrive at the following picture: initially starting out in a resonance, the particle experiences an almost full perihelion distance oscillation; as $q$ begins to decrease, the particle is knocked out of resonance, and wanders in semi-major axis space for some time. Once $q$ begins to increase, the particle lands in a different resonance, and persists in this stable state for the next perihelion distance cycle.

One part of he mystery, however, remains: Why does the resonance hopping occur in the first place? Planet Nine by itself cannot be the reason for the hop; as long as the particle is in resonance, it is generally protected from encounters with Planet Nine, as is clearly shown in Figure \ref{fig:hop}. Once the object exits the resonance, close encounters take place. But if Planet Nine is not driving this complicated behavior, what is the primary mechanism that knocks the bodies out of resonance?

\subsection{Non-Resonant Dynamics} \label{sec:nonres}

Before we consider this problem further, let us address the question of just how representative the object shown in Figure \ref{fig:hop} is as compared to the other anti-aligned objects in the numerical simulations.
{We note that planetary systems are chaotic in nature, with sensitive dependence on initial conditions \citep{Holman1996, Murray1997}. In addition, the orbital elements required to enter into mean motion resonance do not necessarily vary smoothly, but rather occupy fractal regions of parameter space. As a result, descriptions of the dynamics must be presented in terms of the distribution of results (here, for the entire population of test particles). Although not all of the particles will exhibit exactly the same behavior, the description of the previous subsection is representative of the resonance hopping effects explored in this paper}.
Out of the stable anti-aligned objects in our suite of simulated objects, about $51\%$ are in a Planet Nine resonance for at least a subset of the integration time (with the shortest-lived resonances lasting roughly 200 Myr, and the longest-lived resonances lasting the simulation lifetime), a significant fraction of the anti-aligned population. Nonetheless,  a significant number of objects do not experience resonance interactions and remain stable. 

To demonstrate the non-resonant dynamics, Figure \ref{fig:nonres} presents the continuation of the integration for the object shown in Figure \ref{fig:hop}. That is, the red region in the semi-major axis evolution, from about $0.7$ Gyr to $1.1$ Gyr, is the same as shown previously, and corresponds to the $3$:$2$ resonance as discussed above.

Once the object reaches the bottom of its perihelion distance cycle, and is knocked out of the resonance, it fails to attain another resonance once $q$ begins to increase. This is evident in the significant, diffusing variation in the semi-major axis evolution of the object, shown in the top left panel. For the next billion years, the test particle experiences repeated close encounters with Planet Nine (vertical lines), which serve to continually modulate its semi-major axis. Yet, these scattering interactions are not sufficient to cause a violent instability for the object; it is able to remain in the solar system.

Moreover, as we see from the top right panel, the perihelion distance oscillations are hardly affected by the semi-major axis behavior. The object remains anti-aligned (bottom left panel) and the offset of the longitude of perihelion, $\Delta \varpi$, continues to librate about $180^{\circ}$. 

%compute the escape time: two effects: P9 encounters and then Neptune encounters - how many cycles should be able to survive?

\section{Neptune Scattering} \label{sec:scattering}

Let us now return to the original question: In the case of the objects that do experience resonance hopping, what is the mechanism that knocks them out of resonance? In Section \ref{sec:hop}, we noted that the object always exits the resonance as the perihelion distance decreases, at a time when the object is approaching its lowest perihelion $q$ in the cycle. 

The lowest perihelion attained by the particle in Figures \ref{fig:hop} and \ref{fig:nonres}, $q \sim50$ AU, is significantly larger than the orbit of  Neptune (a nearly circular orbit at $a = 30$ AU). A TNO with a perihelion of 50 AU would be classified as ‘detached’ from Neptune by existing classification schemes \citep[e.g.,][]{Lykawka2007, Khain2020, Hamilton2019}. As such, at first glance, it might seem unlikely that a detached TNO could be perturbed enough by Neptune to disrupt the TNO-Planet 9 resonance. In this section, we demonstrate the importance of Neptune even to detached TNOs in this architecture. 

The following analysis is similar in spirit to the Kepler map \citep{Petrosky1986}, first developed to describe the motion of comets in the three body Sun-planet-comet system. The Kepler map can be used to compute kick maps and predict the evolution of solar system comets \citep{Chirikov1989, Malyshkin1999, Rollin2015}. In this work, we use a physically intuitive model to derive the similar effect, where Neptune perturbs the orbits of TNOs in the presence of Planet Nine. 

\begin{figure}
  \begin{center}
      \leavevmode
\includegraphics[width=85mm]{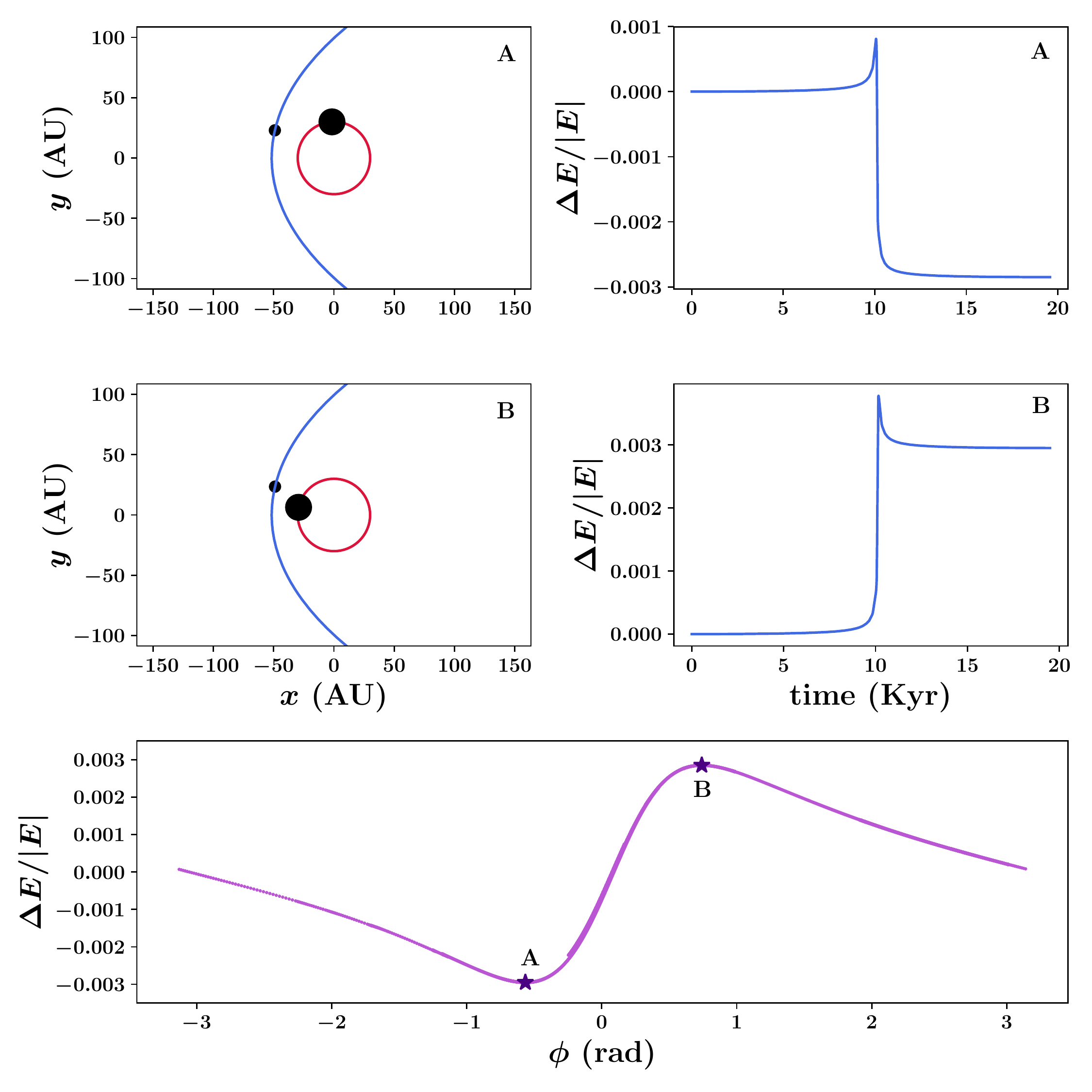}
\caption{Two possible orbital geometries of the scattering interaction between a test particle (on the blue outer orbit with $q = 50, a = 500$ AU) and Neptune (on the red inner orbit with $a = 30$ AU). In the top left panel (A), the force on the test particle from Neptune acts to slow down the the particle. The corresponding loss of energy over one orbit of the test particle ($\sim$12,000 years) is shown in the top right panel. In contrast, in left panel B, Neptune pulls the test particle forward in its orbit, increasing its energy, as is evident in right panel B. The bottom panel shows the change in orbital energy of the test particle for different Neptune positions at the particle's perihelion. A phase of $\phi = 0$ corresponds to Neptune's location being nearest to the particle perihelion position, and $\phi = \pi$, $\phi = -\pi$ corresponds to the location furthest away. The configurations in the top panels (A and B) are labelled on the energy plot with asterisks.}
\label{fig:geometry}
\end{center}
\end{figure}

For any value of the perihelion distance of the test particles, we start by considering the details of their encounters with Neptune. Figure \ref{fig:geometry} displays two possible interaction geometries in the left panels (A and B), where Neptune's orbit is shown as the inner red circle, and the outer blue orbit corresponds to that of the test particle. 

In the top left panel (A), Neptune trails the test particle at the encounter location. As a result, the force from Neptune slows down the particle and thereby reduces its orbital energy. This behavior is illustrated in the top right panel, where the energy increment $\Delta E/|E|$ is plotted over one orbital period of the test particle. The geometry of left panel (B) demonstrates the opposite effect: In this case, Neptune is ahead of the test particle, so that its force acts to speeding it up and increase its orbital energy. 

The bottom panel of Figure \ref{fig:geometry} plots the change in energy of the test particle as a function of Neptune's position at the particle's perihelion, $\phi$ (see \citealt{Pan2004} for an analogous plot for different planet and particle parameters). The variable $\phi = \phi_N - \phi_{TNO}$ can be thought of as a phase difference between the location of Neptune along its orbit ($\phi_N$) and the location of the test particle ($\phi_{TNO} = \pi$). The geometries in the top left panels are labelled on the curve and correspond to the extremal configurations that result in the maximal energy change. This curve is similar in shape to the Kepler maps that can be computed for kicks on (for example) Halley's comet due to the solar system planets (see Figure 2 of \citealt{Chirikov1989} and Figure 2 of \citealt{Rollin2015}), with the exact shape being dependant on the particular parameters of the particle and planet. 

Since the orbital energy of the particle is given by 
\begin{equation}
    E = - \frac{GMm}{2a},
\label{eq:orbitalenergy}
\end{equation}
a change in the energy of a particle results in a change in the semi-major axis, $a$, i.e., 
\begin{equation}
    dE = \frac{GMm}{2a^2}da = -E\frac{da}{a}.
\end{equation}

Figure \ref{fig:geometry} shows that the semimajor axis $a$ can either increase or decrease depending on the arrangement of the test particle and Neptune during the encounter. To understand if this change in semi-major axis is sufficient to drive the resonance hopping effect, we can estimate the size of the energy change from one Neptune interaction by considering an approximate analytical model. 

\subsection{A Random Walk in Semi-Major Axis}
\label{sec:random}

\begin{figure}
  \begin{center}
      \leavevmode
\includegraphics[width=85mm]{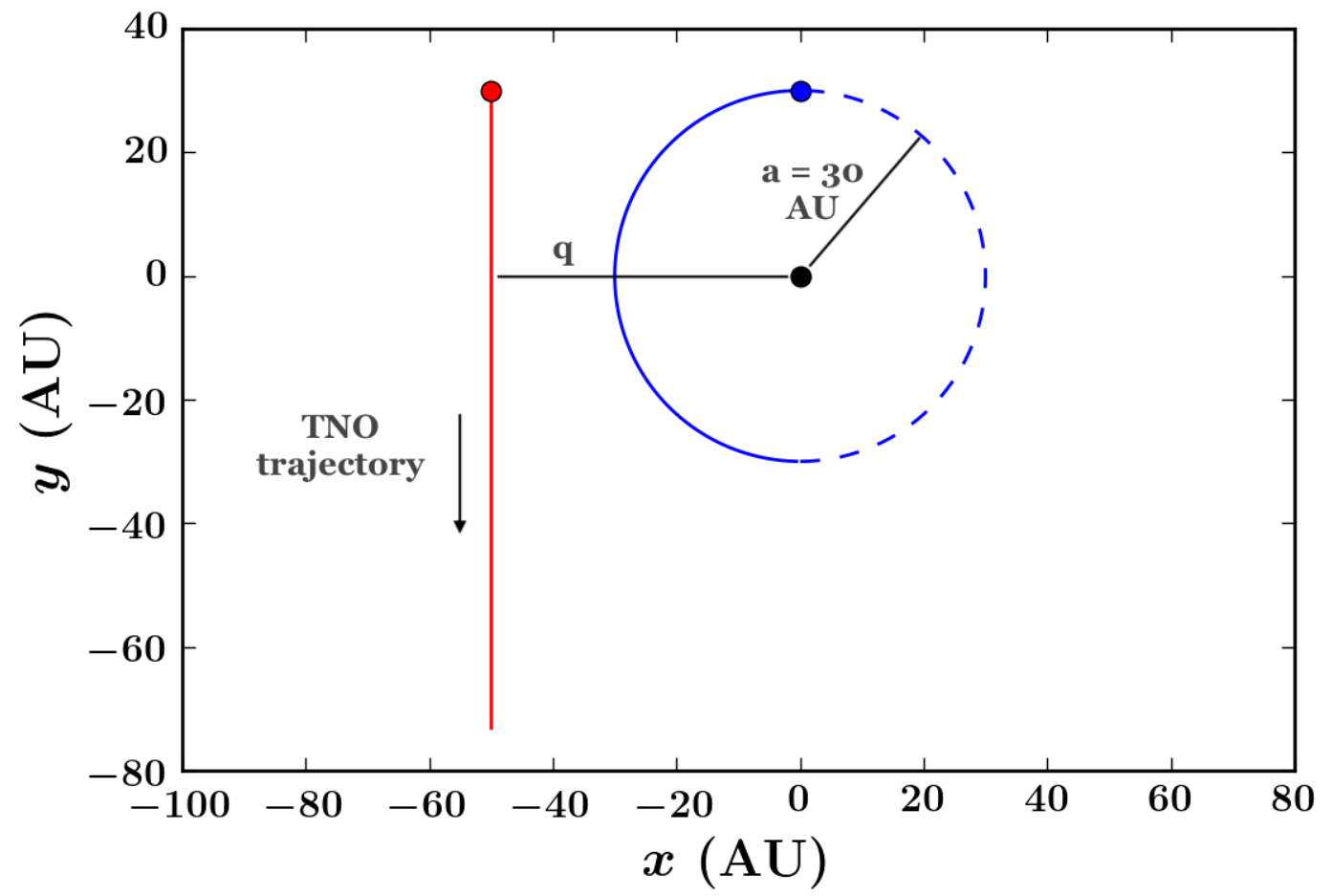}
\caption{The geometric setup for the analytic scattering calculation. The trajectory of the TNO is approximated as a straight line at the particle's perihelion distance ($q$), and the particle travels at a constant speed.}
\label{fig:integral_geometry}
\end{center}
\end{figure}

As discussed above, the energy change for a particular interaction with Neptune is a function of the initial positions of Neptune and the test particle. For simplicity, we construct our model to compute the maximal effect of Neptune on the scattering TNO. The geometry of this extrenal scattering event is a simplification of the the orbital arrangement in Figure \ref{fig:geometry}, and is shown in Figure \ref{fig:integral_geometry}.

We assume that during the encounter with Neptune, the test particle is traveling in a straight line with a constant speed corresponding to its velocity at perihelion.  This heuristic geometry allows us to compute the greatest possible effect that Neptune could have on the TNO during a single encounter. 

In this approximation, the velocity components are given by 
\begin{equation}
    \begin{aligned}
        v_x &= 0\\
        v_y &= v_{peri} = \sqrt{GM (\frac{2}{q} - \frac{1}{a})} \approx \sqrt{ \frac{2GM}{q}},
    \end{aligned}
\end{equation}
where the last equality assumes $a \gg q$.

Now, we would like to compute the energy change $\Delta E/|E|$ of the test particle due to Neptune in this orbital arrangement. Note that
\begin{equation}
\Delta E = \int \vec{F} \cdot \vec{dr} = \int \vec{F} \cdot \frac{\vec{dr}}{\vec{dt}} dt = \int \vec{F} \cdot \vec{v} dt\,.
\end{equation}
Since $v_x = 0$, the expression for  $\Delta E$ reduces to the form 
\begin{equation}
\Delta E = \int F_y v_y dt = v_y \int F_y dt\,.
\end{equation}
\label{eq:deltaE}

The trajectories of the test particle coordinate $(x, y)$ and those of Neptune $(x_N, y_N)$ are given by
\begin{equation}
    \begin{aligned}
        x(t) &= -q\\
        y(t) &= v_y t + a_N\\
        x_N(t) &= a_N \cos{(\frac{2\pi}{T_N}t + \frac{\pi}{2})}\\
        y_N(t) &= a_N \sin{(\frac{2\pi}{T_N}t + \frac{\pi}{2})}\\
    \end{aligned}
\end{equation}
where $a_N$ is the semi-major axis of Neptune, and $T_N$ is Neptune's orbital period.

The gravitational force between Neptune and the test particle is
\begin{equation}
\vec{F} = \frac{GM_Nm}{r^2} \vec{\hat{r}}\,.
\end{equation}
As a result, 
\begin{equation}
F_y = F \frac{r_y}{r}\,,
\end{equation}
where $r_y = |y(t) - y_N(t)|$, the vertical distance between the test particle and Neptune.

In this geometry, we assume that the interaction between Neptune and the test particle occurs over half of Neptune's orbital period. Integrating Equation (\ref{eq:deltaE}) from 0 to $T_N/2$, for $q = 50$ and $a = 500$ AU, we find that 
\begin{equation}
    \Delta E/E \approx 0.0023.
\end{equation}
We can compare this value to the one shown in the right panels of Figure \ref{fig:geometry}, which shows numerical simulations of an analogous, but realistic, geometry. The agreement between the model and simulations suggests that this approximation accurately predicts the upper bound on Neptune's effect on the TNO during one encounter. 

In the context of distant TNOs that interact with Planet Nine, the perihelion distance of the test particle remains near its minimum for millions of years (Figure \ref{fig:hop}). As a result, the TNO interacts with Neptune every orbit for an extended interval of time. Since the orbital period of the test particle in question is much greater than that of Neptune, Neptune's locations at each perihelion crossing of the particle are uncorrelated\footnote{This holds true as long as the test particle is not in a mean-motion resonance with Neptune. The most distant object known to be in resonance with Neptune resides at roughly $\sim130$ AU \citep{2018AJ....155..260V}, and since the objects we are considering in this work have $a > 150$ AU, this is a reasonable assumption to make. In the absence of Planet Nine, mean motion resonances with Neptune can extend to larger semi-major axes \citep[$a~>~150$ AU; see ][]{Saillenfest2020}; {however, in the presence of Planet Nine, TNOs with longer period orbits will likely fall into lower order resonances with Planet Nine than with Neptune. A consideration of the relative expected population of the most distant Neptune resonances in the presence of Planet Nine would be an interesting avenue for future study.}}. As a result, when the test particle approaches perihelion (and Neptune's orbit), there are many possible scattering geometries, each with its own encounter strength. 

Since the semi-major axis can increase and decrease due to these interactions, we expect to see a diffusion process in the semi-major axis over many orbits \citep[as shown in][]{1987AJ.....94.1330D}. To test this idea, we turn to a suite of numerical simulations.

\vspace{-0.7mm}

\subsection{Numerical Scattering Experiments}
\label{sec:numscat} 

To test whether the test particles experience diffusion in semi-major axis due to interactions with Neptune, we run a number of Neptune scattering experiments. Using the \texttt{mercury6} integrator as before, we now include an active Jupiter, Saturn, Uranus, and Neptune, and exclude Planet Nine. The orbits of the test particles are initialized over a range of semi-major axes ($a = 150 - 700$ AU) and perihelion distances ($q = 40-50$ AU). All of the angular orbital elements (inclination, longitude of ascending node, argument of perihelion) are taken to be zero. The simulations contain one hundred particles for each pair of ($a$, $q$) values, with varying mean anomalies.

\begin{figure}
  \begin{center}
      \leavevmode
\includegraphics[width=85mm]{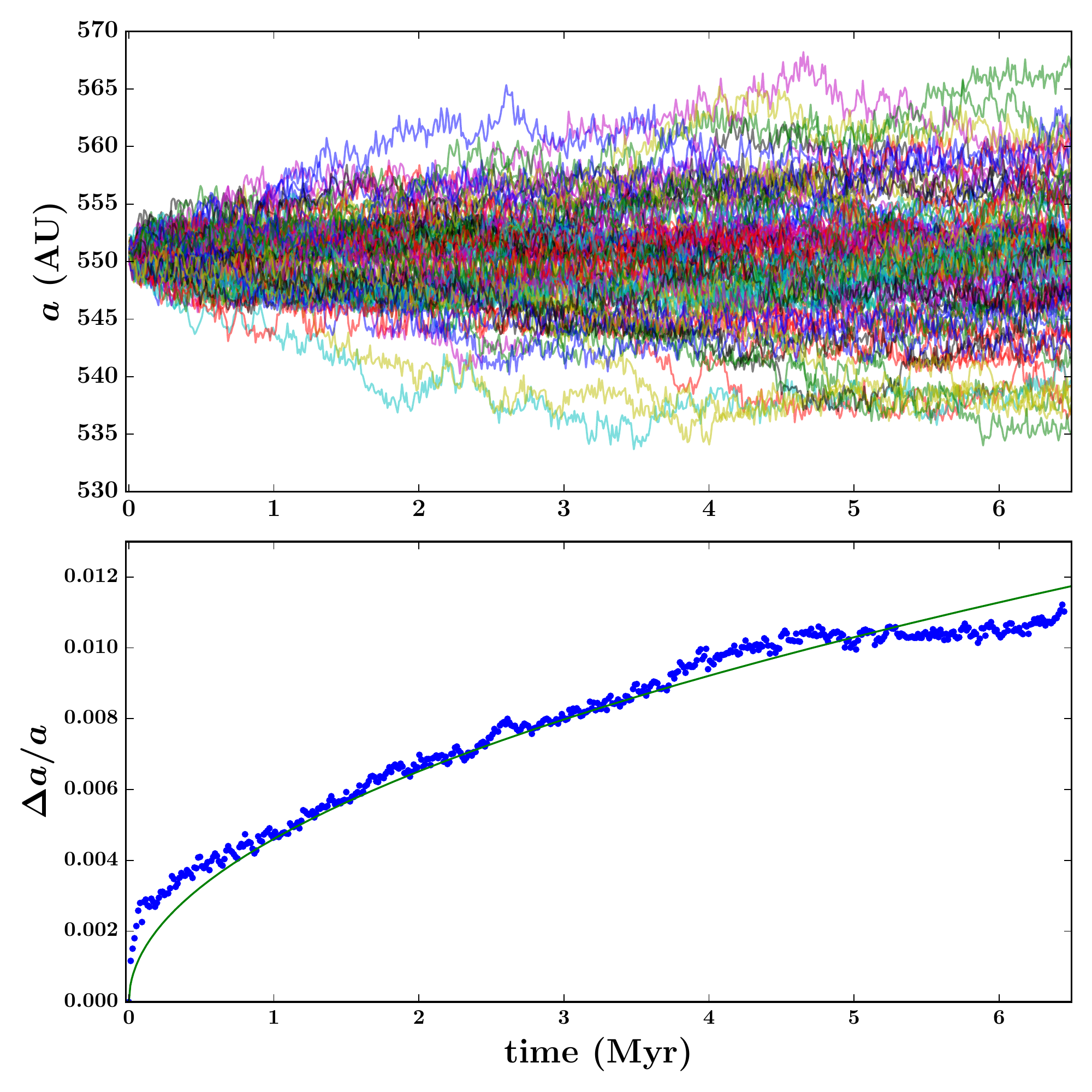}
\caption{Semi-major axis diffusion of a set of test particles. The top panel shows the $a$ evolution of one hundred test particles initialized at $a = 550, q = 50$ AU, with varying initial mean anomalies. The growth of the semi-major axis distribution is fit to a pure diffusive process in the bottom panel. Here, $\Delta a = \sqrt{\langle a^2 \rangle}$.}
\label{fig:diffusion}
\end{center}
\end{figure}

An example of the integration results is shown in Figure \ref{fig:diffusion}. The top panel shows the semi-major axis evolution of one hundred particles initialized with $a = 550$ AU, $q = 50$ AU over 500 orbits. Note that the semi-major axes of these particles change on average by more than a few AU over less than 1 Myr. In this high-$a$ regime, it appears that relatively large perihelion distances do not protect objects from Neptune scattering. These significant changes in semi-major axis are sufficient to knock a particle out of resonance. As shown in Figure \ref{fig:diffusion}, after some time ($\sim$1 - 10 Myr), the accumulated changes in semi-major axis will become substantially greater than a few AU, the typical width of a resonance (e.g. \citealt{2017AJ....154...20W}). These results indicate that Neptune could indeed drive TNOs out of resonance with Planet Nine and thus instigate the resonance hopping effect. 

Over a few Myr, the spread in the semi-major axis distribution significantly increases, while the perihelion distance of the particles does not change. To quantify this spread, we compute the mean squared change in semi-major axis for all one hundred objects, and divide by the initial semi-major axis.
\begin{equation}
\frac{\Delta a}{a} = \frac{\sqrt{\langle a^2\rangle}}{a} = \frac{1}{a}\sqrt{\frac{1}{N}\sum{\left(a_i(t) - a_i(0)\right)^2}} \,.
\end{equation}

The bottom panel of Figure \ref{fig:diffusion} plots the evolution of the semimajor axes due to the increments $\Delta a/a$ outlined above. We find that the spread from the top panel can indeed be described by a diffusion process: $\Delta a/a$ increases as the square root of time (or more precisely, as the square root of the number of orbits, equivalently, interactions). As a result, we can consider the orbital evolution of the test particles to be a random walk in semi-major axis.

\begin{figure}
  \begin{center}
      \leavevmode
\includegraphics[width=85mm]{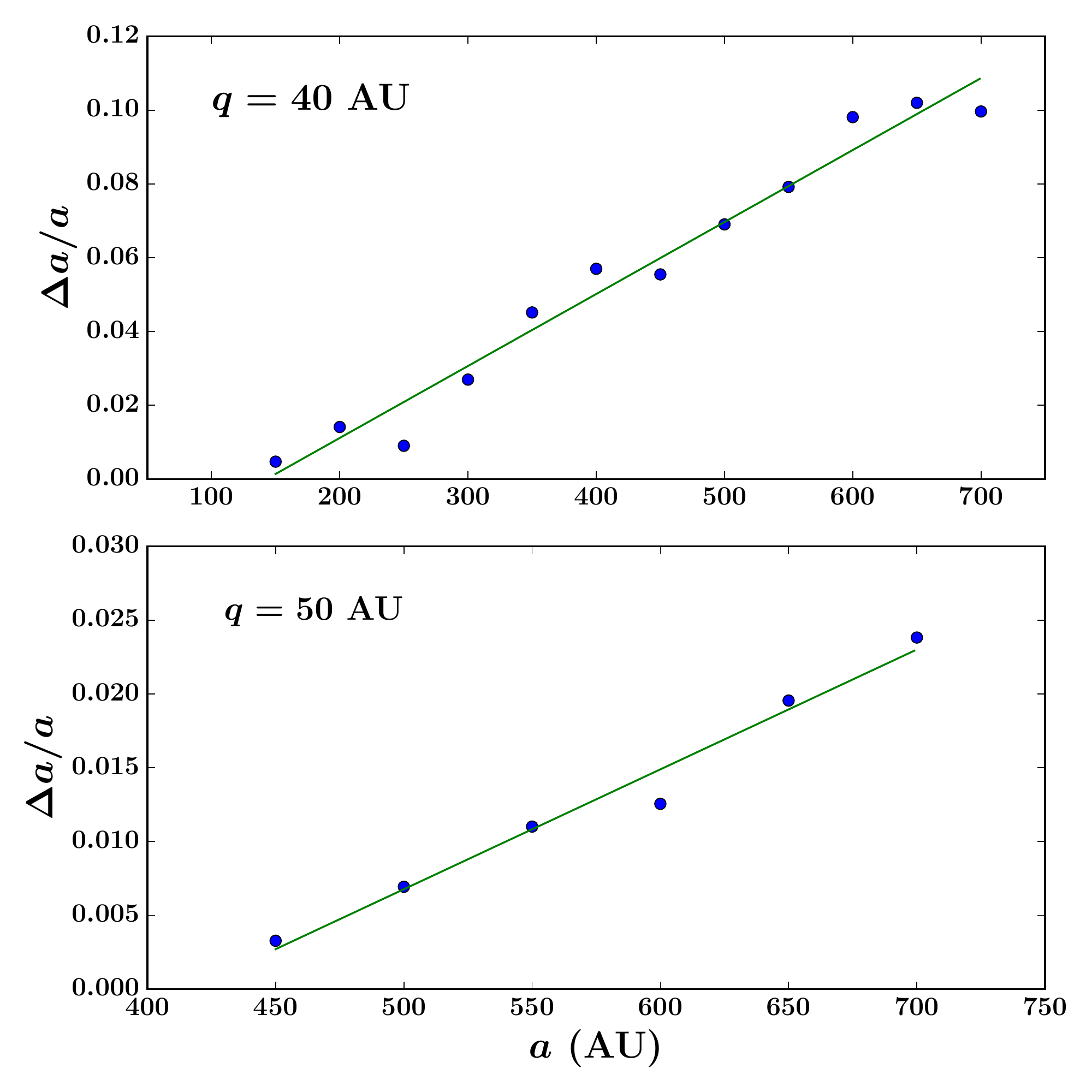}
\caption{Dependence of $\Delta a/a$ on semi-major axis for $q = 40$ AU (top panel) and $q = 50$ AU (bottom panel) from numerical scattering experiments. Each marker denotes the mean squared change in semi-major axis for all one hundred objects over a time interval of five hundred orbits, as in Figure \ref{fig:diffusion}. Both plots show a linear relationship, as predicted by the simplified model.}
\label{fig:adep}
\end{center}
\end{figure}

We repeat the analysis shown in Figure \ref{fig:diffusion} for a set of initial $(q, a)$ pairs. The results are illustrated in Figure \ref{fig:adep}: The top panel plots the dependence of $\Delta a/a$ on the initial $a$ for $q = 40$ AU and the bottom panel plots the same quantities for $q = 50$ AU. We find that not all regions of parameter space demonstrate the diffusion process described above. For $q = 40$ AU, all objects initialized with $a \lesssim 150$ AU evolve with nearly constant semi-major axis. Similarly, objects with $q = 50$ AU and $a \lesssim 450$ AU are not affected by Neptune scattering.

In both $q = 40, 50$ cases, we find that objects with larger semi-major axes are more strongly affected by Neptune interactions, as $\Delta a/a$ increases with $a$. Intuitively, this trend occurs because objects with larger $a$ are less bounded to the system (Equation \ref{eq:orbitalenergy}), and thus experience greater relative energy changes due to the same Neptune kick.

In addition, we find that the distribution of semi-major axis changes over one orbit of a set of test particles with random initial conditions is equivalent to the distribution of semi-major axis changes over many orbits of the same particle. This result confirms our assumption that Neptune's locations at each perihelion crossing of the test particle are uncorrelated for $a \gg a_N$.

Interestingly, our analytic model predicts that $\Delta E/E$ (or $\Delta a/a$) is proportional to $a$, as confirmed by our numerical simulations in Figure \ref{fig:adep}. However, further comparisons are difficult to make for a few reasons. First, our model is valid in the limit $a \gg q$, and thus fails to predict the correct y-intercept for Figure \ref{fig:adep}. In addition, the model only computes the maximum change in energy over one encounter, and thus overestimates the size of the true random walk step. For example, multiplying the relative energy change predicted by the model for $a = 500$, $q = 50$, by the number of orbits in Figure \ref{fig:adep}, we find
$0.0023 \sqrt{500} \approx 0.051,$
a value much greater than the corresponding result $\Delta a/a = 0.008$ shown in the bottom panel of Figure \ref{fig:adep}. 

Finally, we note that the distribution of semi-major axis changes at each perihelion crossing may be non-trivial. This complication prevents us from easily comparing the true random walk in simulations with the one predicted by the model. 

%Additionally, note that $\Delta E$ is only a function of perihelion distance, $q$. Thus, the only dependence on semi-major axis, $a$, comes from the orbital energy, $E$.

%That is, 
%\begin{equation}
%    \frac{\Delta E}{E} \propto a.
%\end{equation}

%According to this simplified model, the larger the semi-major axis of the test particle, the stronger the effect of Neptune on its orbital energy. In Figure \ref{fig:adep}, we compare this relationship to our numerical scattering simulations. We plot the $\Delta a/a$ dependence on semi-major axis for two values of perihelion distance, $q = 40, 50$ AU. The simulations clearly follow a linear relationship, as predicted by the model. %In addition, the theory allows us to compute the slope, $m$, of the linear fit:

%\begin{equation}
%    c = \frac{2 v_y \int F_y dt}{GMm}
%\end{equation}

%Figure \ref{fig:adep} shows the comparison between the slope predicted by the model and the measured value in the simulations. Both the $q = 40$ and $q = 50$ cases show reasonable agreement with the simplified model, confirming the underlying scattering mechanism.

Nonetheless, the analytic model allows us to understand the numerical results as a Neptune-driven diffusion process. Combining the results presented above, we find that interactions with Neptune affect the distant ETNOs, even at large perihelion distances and particularly at large semi-major axes. Specifically, Neptune's influence can be modeled as a random walk in either energy or semi-major axis, where each step of the walk occurs once an orbit of the test particle. In a fairly short amount of time (on the order of 1 Myr), the accumulated change in semi-major axis of the test particle exceeds the typical resonance width, knocking the particle out of the resonant stable state with Planet Nine. As a result, even the distant Neptune scattering interactions are in fact sufficiently strong to induce the resonance hopping mechanism.

\vspace{-0.5mm}

\section{Discussion} \label{sec:disc}
 
With the results presented above, we can now summarize the resonance hopping effect in the Planet Nine-Neptune system. In numerical simulations with Planet Nine, we find that a subset of the anti-aligned test particles are in fact in mean-motion resonances with Planet Nine. These objects are able to reside in their resonant states for time spans comparable to the period for oscillations of the perihelion distance, where this time scale is of order half a billion years. 

As the test particles approach smaller perihelion distance values ($q \sim 50$ AU), the Neptune scattering interactions perturb the resonance dynamics and thereby knock the particles out of resonance. Without the phase protection mechanism, the particles can now experience close encounters with Planet Nine, in addition to scattering effects from Neptune, and these two types of interaction result in significant variations in the semi-major axis. As the perihelion distance begins to increase, as the secular cycle continues, the objects may or may not land in another resonance.

If the test particle does enter a new resonance, it is guaranteed enhanced stability over the next cycle in perihelion distance. With a resonant configuration, the body is protected from Planet Nine encounters and is sufficiently detached from Neptune so that scattering effects are weak. 

If, however, the object does not land in a resonance, it may still survive. The secular cycles in perihelion distance and the longitude of perihelion are not affected by changes in semi-major axis. As long as the close encounters with Planet Nine do not significantly disrupt the orbital evolution, the test particle can expect to survive over some number of secular cycles. The secular time scale for these TNOs is on the order of $\sim1$ Gyr; as a result, the age of the solar system ($\sim4$ Gyr) does not provide sufficient time for these objects to become unstable. Computing the expected lifetime of these objects, taking into account both Neptune and Planet Nine scattering events, would be an interesting avenue for future work.

In Section \ref{sec:scattering}, we show that the Neptune scattering events that lead to the resonance hopping effect can be modeled as a diffusion process in energy or semi-major axis. We present simplified numerical and analytical models that successfully estimate the largest possible step in this random walk. In the real system, however, the energy step size is not the same each time, as the encounter strength depends on the location of Neptune as the test particle approaches perihelion. As a result, the Neptune scattering interactions are better modeled as a random walk with varying step size, where the step size is drawn from some underlying distribution. Future work should measure this distribution in simulations and/or analytically determine its form, and then use the results to construct the adjusted diffusion process. 

\begin{figure}
  \begin{center}
      \leavevmode
\includegraphics[width=85mm]{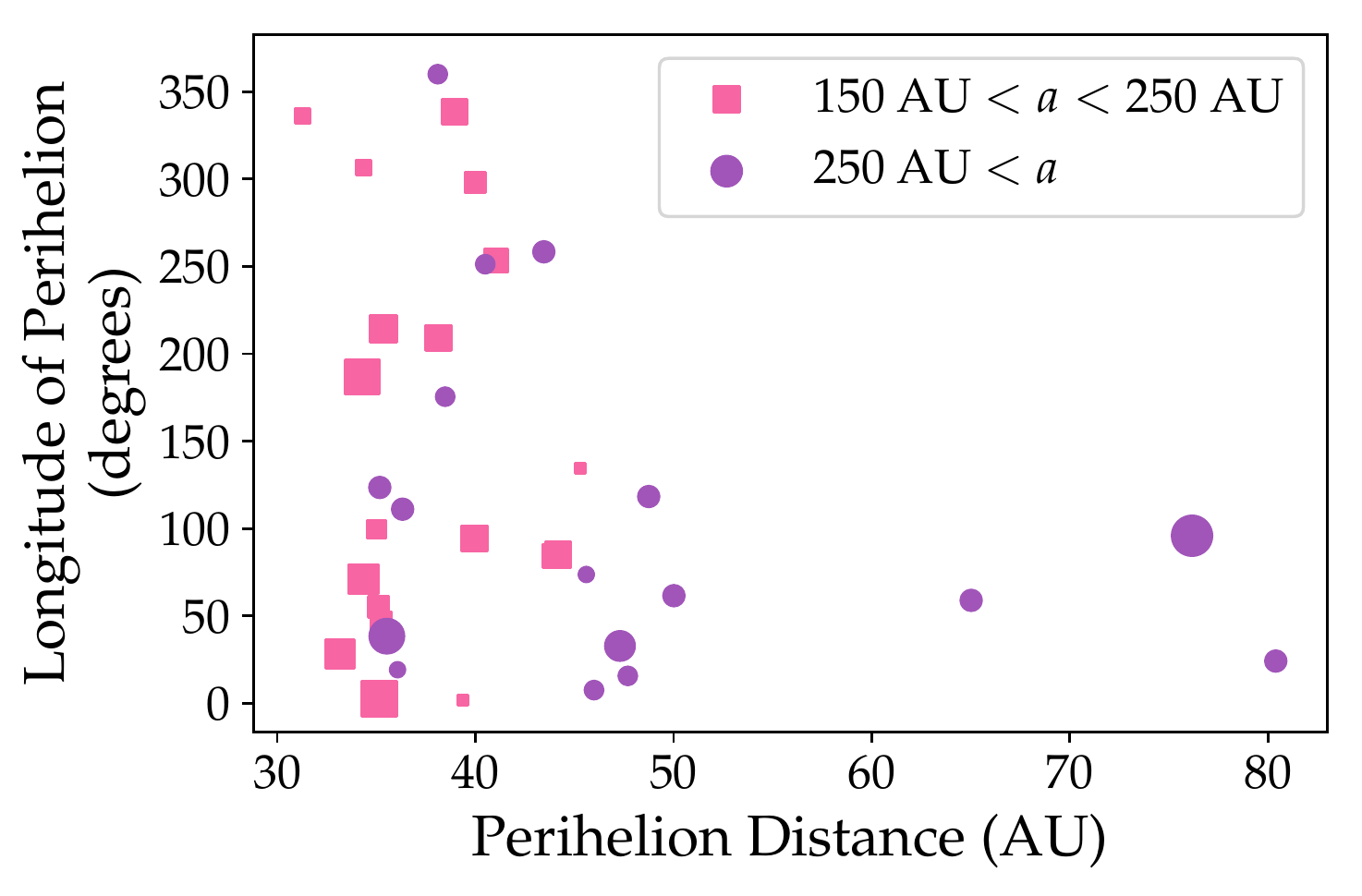}
\caption{The measured longitude of perihelion ($\varpi = \omega + \Omega$) for the discovered TNOs with semi-major axis greater than 150 AU, plotted by perihelion distance. As discussed in \citet{P9Review}, the range in longitude of perihelion values decreases with larger perihelion, a manifestation of Neptune interactions being less important on the more distant TNOs (as shown in Figure \ref{fig:adep}). The points are sized by the number of oppositions used to compute the orbit {(which ranges between 1 and 15)}, with larger points corresponding to longer arc lengths. The three points with the three largest perihelion values are 2015 TG387 \citep[$q \approx 65$ AU;][]{Sheppard2019}, Sedna \citep[$q \approx 76$ AU;][]{Brown2004}, and 2012 VP113 \citep[$q \approx 80$ AU;][]{2014Natur.507..471T}. Data was retrieved from the Minor Planets Center (\url{https://minorplanetcenter.net/}) on 8/25/2020 and no data quality cuts were made.}
\label{fig:pericluster}
\end{center}
\end{figure}

In Figure \ref{fig:pericluster}, we show the distribution of longitude of perihelia for the known sample of TNOs with semi-major axes $a> 150$ AU. {Although the number of known objects with particularly large perihelion distances is very low, the width of the distribution appears to decrease with increasing perihelion distance.} In other words, the cluster of orbits identified as evidence for Planet Nine becomes more pronounced as perihelion distance increases and the importance of Neptune interactions decreases \citep[see the discussion of this observation in][]{P9Review}. The TNO orbits that are primarily dominated by secular effects spend less time in the resonance hopping regime, which depends on perihelion distance (as found in this work); these `secular-dominated' TNO orbits are thus more likely to reside in the cluster identified in \citet{2016AJ....151...22B}. {Although the observational sample is currently too small to make any definitive conclusions at the present time, the} relationship between TNO perihelia and cluster width is another fruitful avenue for future study, particularly when combined with the underlying distribution of Neptune interactions expected at each distance. 

%questions to think about: more statistics about ejections based on semi-major axis/perihelion distance? destabilization always at the lower end of the perihelion cycle?

Further confirmation of the importance of Neptune in the resonance hopping effect comes from the work of \citet{2018AJ....156...74B}. This study considers systems in which Planet Nine is the only massive perturber and the giant planets --- Jupiter through Neptune --- are accounted for with a $J_2$ approximation. In these simulations, the anti-aligned objects remain in the same resonance for the entire length of the integration time. The lack of resonance hopping in this case is due to the absence of an active Neptune, and consequently, the absence of TNO-Neptune scattering interactions. Similarly, work in \citet{2017AJ....154...61B}, which treated only Neptune and Planet Nine as active particles, found that dynamically stable objects can change resonances many times during integrations of the same length. While \citet{2017AJ....154...61B} did not consider whether objects were aligned or anti-aligned with Planet Nine, the increased rate of movement between resonances compared with the case where Neptune is treated as a perturbation on the solar potential is consistent with our current work. 

In this work, we use the Planet Nine orbit from the initial \citet{2016AJ....151...22B} study, with $a_9 = 700$ AU and $e_9 = 0.6$. Although recent estimates have favored a somewhat lower semi-major axis and lower eccentricity orbit for Planet Nine \citep{P9Review}, we choose to retain the original proposed orbit for ease in making comparisons with earlier work. The effect of Planet Nine on the outer solar system is quite robust: all of the orbits in the generally accepted range of parameter space (see Figure 15 of \citealt{P9Review}) achieve a qualitatively similar structure, and thus the exact orbit does not matter as long we view the results through a statistical lens. This is shown in Figure 17, 18 and 19 of \citet{P9Review}, which recreates the phase space trajectories for three distinct orbits of Planet Nine, and the behavior in all three phase spaces is similar (although the exact parameters of TNOs that will appear clustered may change). 

Following previous studies, we have considered this resonance hopping situation in the near planar case, with slightly inclined test particles but a non-inclined Planet Nine. Even in this idealized context, it is clear that residing in a resonance is not necessary to maintain anti-alignment nor long term stability. The next step in continued analysis of this problem is to consider resonance hopping effects in fully inclined simulations. Although we expect to find fewer resonant events in this more realistic situation, the relation between the secular cycles in perihelion distance and the Neptune scattering interactions is likely robust and is expected to yield qualitatively similar results.

\vspace{-0.5mm}

\section{Conclusion} \label{sec:conclude}

The goal of this work was to explore the dynamics of TNO interactions with both Neptune and a hypothetical Planet Nine, where the TNO orbits transition from one resonant configuration with Planet Nine to another (this phenomenon is sometimes known as resonance hopping). In this context, the dynamics take place in the outer solar system, where the TNO orbits have long periods and high eccentricities. This work is thus a variation of the classical problem concerning the dynamics of high eccentricity comets that trace through the inner solar system (where the evolution can be described using the Kuiper map). In addition to the implications for the Planet Nine Hypothesis, the main result of this work is a more detailed description of the resonance hopping effect, where our specific findings can be summarized as follows:  

$\,$ 

\medskip
\noindent
[1] In the presence of Planet Nine, the perihelion distance $q$ of the TNO orbits executes secular variations with time scales of $\sim1$ Gyr, with corresponding cycles of the orbital elements $(\Delta\varpi,e)$. The relative importance of interactions with Neptune and Planet Nine depend sensitively on the value of $q$ (as illustrated in Figure \ref{fig:hop}). The TNO orbits remain in resonant states during the large-$q$ portion of the secular cycles, but tend to leave resonance when $q$ decreases. 

\medskip 
\noindent 
[2] The proximity of the TNO orbit to Neptune leads to scattering interactions that act to remove the TNO from resonance with Planet Nine, for values of the TNO perihelion as large as 50 AU. These Neptune-scattering interactions are effective, in spite of the large distances between the bodies, because the distant TNO orbits have relatively small binding energies. For fixed perihelion, the effectiveness of Neptune scattering increases with semi-major axis of the TNO (Figure \ref{fig:adep}). 

\medskip
\noindent
[3] The scattering interactions with Neptune lead to changes in the semi-major axes (equivalently, energy) of the TNO orbits that can be described by a random walk (Figure \ref{fig:diffusion}). The random walk (in $a$) allows some TNOs to enter a new resonance with Planet Nine (resulting in successful resonance hopping), whereas the semi-major axes of other TNOs continue to evolve (Figure \ref{fig:nonres}). We have constructed a simple analytic model to account for this behavior (Section \ref{sec:random}), which provides an upper limit for the increment of energy changes (see also \citealt{Pan2004}). Numerical simulations of these scattering interactions (Section \ref{sec:numscat}) provide qualitatively similar results, but include more complicated behavior, and indicate that the variations in semi-major (energy) sample a distribution of values. 

%As cycles in the TNOs' perihelion distances (in $e-\Delta \varpi$ space) arise naturally due to secular interactions with Planet Nine, the times at which a TNO is most likely to hop between Planet Nine resonances can be determined by its  proximity to Neptune (measured by its perihelion distance). 
    
\medskip 
\noindent 
[4] Over most of the age of the solar system, the most distant TNOs, particularly those with $a>250$ AU, are predicted to reside in regimes of parameter space where their primary interactions with other bodies are secular in nature. Significantly, even the scattering interactions that perturb the semi-major axis (for TNOs near perihelion) do not disrupt the secular cycles. 

\medskip
\noindent
[5] The extreme TNOs that provide evidence for the existence of Planet Nine can remain in stable orbits in the absence of resonances. This finding indicates that the dynamics of the extreme TNOs are primarily influenced by secular --- rather than resonant --- interactions with Planet Nine. For observed TNOs, the spread in longitude of perihelion $\varpi$ decreases with increasing $q$ (Figure \ref{fig:pericluster}). This result is consistent with more distant objects being less influenced by Neptune scattering interactions. 

%Due to the $e-\Delta \varpi$ cycles experienced by TNOs in the Neptune-TNO-Planet Nine geometry, for a large fraction of the TNO's orbit, it resides in a regime where its primary interactions with other bodies in the system are secular in nature. 

\medskip
{Acknowledgements.} We thank Konstantin Batygin for useful discussions and suggestions. The computations for this work used the Extreme Science and Engineering Discovery Environment (XSEDE), which is supported by National Science Foundation grant number ACI-1548562. This research was done using resources provided by the Open Science Grid, which is supported by the National Science Foundation and the U.S. Department of Energy's Office of Science, through allocation number TG-AST190031. J.C.B.~has been supported by the Heising-Simons \textit{51 Pegasi b} postdoctoral fellowship.

Software: pandas \citep{ mckinney-proc-scipy-2010}, IPython \citep{PER-GRA:2007}, matplotlib \citep{Hunter:2007}, scipy \citep{scipy}, numpy \citep{oliphant-2006-guide}, Jupyter \citep{Kluyver:2016aa}

\bibliographystyle{mnras}

\end{document}